# Unit verification procedure as a test of real time messaging-based processes

Miklos Taliga , PhD student
Budapest University of Technology and Economics
Department of Control Engineering and Information Technology
Budapest, Hungary
www.iit.bme.hu, taliga@iit.bme.hu

*Abstract*—The article presents the first results of a PhD study connected to testing of safety critical medical devices: a systematically executed case study at a Hungarian manufacturer of medical devices. . The article shortly describes the process of testing currently being used. Elements of the testing approach less commonly applied in software industry are emphasized . The ending point of the actual testing process in the case study is the starting point for further research: the automated analysis of the testing results. The author started to develop a new approach, using a combination of tools, and modeling a model-based test generating tool – something that is both novel and intensive as an area of research.

## I. Introduction

The concept of „appropriate testing" is not yet well defined; there are multiple schools with multiple – mostly theoretical – models for the „best" testing techniques and approaches. However, the testing approach used is – ultimately - a choice of the developer and the customer of the software, and it largely depends on the business context, as well as the theoretical knowledge about software quality attributes, and the ability to combine them. Within our PhD research, our first aim is to analyze the problems faced by manufacturers involved in development and testing of safety-critical software. We concentrate primarily on the quality attributes connected to security, safety and user trust, seeking understanding about the definition, decomposition and measurements of these mostly theoretical concepts. Next, we try to give possible solutions for the problems found. Our belief is that a solid theoretical framework is needed in order to show how these concepts - frequently used in testing but not fully understood – interrelate. There is indication that a set of important quality attributes can be identified for different safety-critical situations, and a set of guidelines, containing best practices about testing such systems can be built. Finally, we aim to develop a model supported by one or more software testing tools to promote security, reliability and trust.

## II. THE CASE STUDY

### A. The environment and the product

A medical device developer company – where the author has a background in testing – agreed to host our first case study. This company not only sells, but also operates and maintains its products. A fortunate situation, since the customer provides immediate feedback that affects development and testing directly. The device currently in development is an *acute dialysis machine* that will be used in the intensive care units of major hospitals and dialysis centres. The development of the acute dialysis machine utilises the developmental and operative experience in chronic dialysis machines accumulated in the past decades.

### B. The testing process and standards used

Safety, security and user trust are all extremely important aspects if the production and operation of the dialysis machine. The company concentrates strongly on systematic development and testing. It is worth mentioning that this company is among the first ones worldwide using Medical SPICE (see http://medispice.ning.com/) to improve the capability of their processes. The development and testing process are organized around the V-model for medical devices, that is found in 62304 standard (see [2]). The approach used at the company contains steps for the development of medical electrical device or electrical system (PEMS - Programmable Electrical Medical System) and the programmable electronic subsystem (PESS - Programmable electronic subsystem). Within this article we focus on a software-based verification method, which could be highlighted in the "bottom" of the V-model. The testing process used at the company is the following: based on the specification, test scripts are developed; next, "units" (hardware and software components) to be tested are identified, and the "tasks" executed by them as well. Test environment is generated for each task. The test environment simulates the communication of all tasks around the "Task Under Test" – TUT. A test program was developed with the contribution of the author; this is used to do the dynamic testing based on the specification, in an automated way. As the timeframe of the test can be measured in milliseconds, the test program is run many times. A log is created, containing large number of items. The analysis of this log is extremely important to decide whether the tests were successful or not. The solution being used by the company to do this analysis is subject of III. However, it is worth to mention already at this point that the actual solution is much more than a commonly used approach, and the company is investing in further research to improve the automated analysis of test results (see IV). During the development of medical equipment, the need arises to generate objective results, measure parameters imperceptible to humans, acquire multiple forms of information, and also for supported evaluation, data storage and controlled therapy.

| Name | Standard | Description | Effect on medical equipment or standard |
|------|----------|-------------|------------------------------------------|
| Medical equipment management standards | ISO14971 ISO13485 | Designates the basics of medical equipment development | Influences the development of medical equipment |
| Medical equipment process standard | IEC 62304 | Provides a detailed guide to the development and maintenance of a secure software system | These standards appear as an input requirement for medical equipment management standards, as they form the basis of these standards these standards also influence the development of medical equipment |
| Medical equipment product standards | IEC 60601-1 | Provides specific guidance to the production of safe medical equipment | These standards influence the realization of medical equipment management standards, and indirectly affect the development of medical equipment |
| Miscellaneous standards | IEC/ISO 12207 IEC 61508-3 IEC/ISO 90003 | Supplementary guidelines, techniques, etc. that may be useful during development | These support the development of medical equipment |

**Table 1. : Standards affecting the development of medical equipment. See [2].**

When such medical equipment is to be developed, the final product must comply with the regulations of relevant directives which also serve as a kind of integrated certification. To prove compliance, the product receives a CE notation which must be clearly marked. Directives are distinguished by whether the developed equipment will be used inside the human body or as an in vitro diagnostic tool. From the perspective of medical equipment, it is a device that, as per the definition of the manufacturer, performs in vitro examinations on samples (blood or tissue) and produces results thereafter. MDD (Medical Device Directives – 93/42/EEC) denotes European directives that define basic concepts, requirements, how medical equipment enters the market, the method of use, the establishment of technical documentation or design documentation and the system of supervision and quality assurance. Directives are more than guidelines, as they legally oblige in all EU member states. Medical equipment that does not comply with these legal obligations may not enter the market and thus will not be used. The realizations of medical equipment directives must be monitored through the use of examinations based on harmonized standards. During the development of software for medical equipment, some standards determine the basics of development, others offer detailed guidelines to the development and maintenance of a safe software system, yet others provide specific guidelines supplementary guidelines and techniques that may prove useful during the process of development. Table 1 summarizes the standards that contribute to the development of safe medical equipment.

## III. Actual problems and possible solutions

The ending point of the actual testing process in the case study is the starting point for further research: the automated analysis of the testing results. At this moment a static tool is being used to assess testing efficiency (using only coverage metrics). The next figure shows that there are additional software system components between the hardware elements of the PEMS and the Graphical User Interface (Figure 1) Requirements of the software unit test in the IEC 62304 standard can be fulfilled by using the Control System (CSS), Display System (DSS) and the Protective System (PSS). The internal representations of various hardware units (sensors, actuators) are established in the CSS and PSS (in an embedded microcontroller architecture).

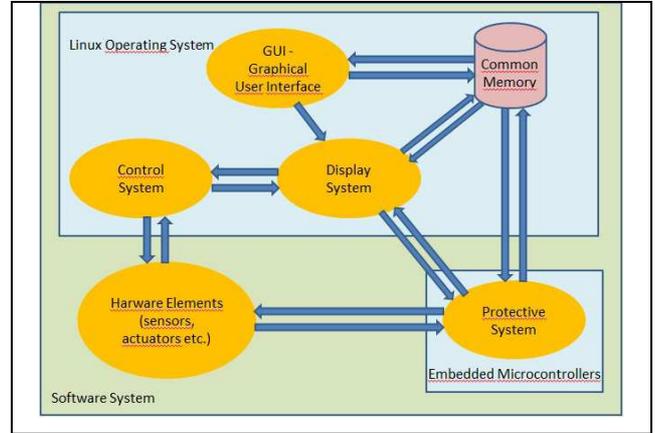

**Figure 1 : Generated task environment**

The DSS (Display System) creates contact between the CSS, DSS and the Graphical User Interface. In this case, the DSS is comprised of tasks and the tasks communicate via messages sent through system calls. The message content is a data structure which contains unique information concerning the task. Communication between the interfaces is recorded when the DSS unit test is running. After processing the log file made from the messages, the test result is displayed – keeping new technologies in mind – in a browser after HTML parsing (Figure 2, Figure 3). For instance: information during the test of a hardware component (e.g.: a sensor) starts from the hardware component. It is then processed by the Control System and the result is sent to the Display System. The Display System stores information as a data structure on the Common Memory (CM). Finally, the Graphical User Interface extracts this information from there and displays it for the user. Communication between the tasks may be achieved through the use of a timer task or a message triggered function.

**Figure 2 : Log file**

As this medical device belongs to the family of embedded devices, thy system is also control-oriented. Moreover, as functions of control-orientation may be described simply as a response to an event in the outside world, these functions can be modelled accurately by a state machine or a state chart. In DSS, communications between interfaces denote functions that are possible to describe with state machine as well as a state chart.

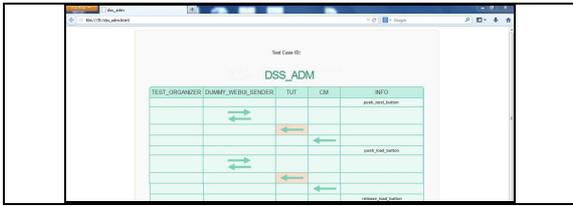
**Figure 3: Results of DSS unit test displayed after HTML parsing**

A sufficiently robust DSS with a capacity for lots of functions may be described with a state chart. One may ask how new functions are added to the system and to what extent does that change the entire model? How open must this model be designed so that new information does not change it fundamentally, necessitating a redesign. One possible answer is that state diagrams for state machines must be designed with care. As this method is not an iterative design method and there is no the levels of hierarchy cannot be supervised, a mapping system that eliminates these shortcomings is necessary for the sake of expedience. Such a system is, for example, the UMLS03 standard, that contains a state chart diagram kit for the design of stat charts. Naturally, a state space designed so must still be checked. The complete state space of the model describing the behaviour of the system must be explored and every state must be examined. Certain model checking tools already exist to this end, that are low level, mathematically well manageable formalisms (e.g.: Kripke structure or an LTS – Labelled Transition System). The resulting complete state chart must then be transformed to a language that is accessible to the model checking tool. The model checker must also automatically generate test cases in accordance with the provided specifications. Obviously methods for examining the code line coverage must also be written at this point to achieve a complete verification process.

## IV. Future development

The need may arise to connect the output of the unit test to a tool capable of running the test automatically. Such a tool would be the free, open source program, Jenkins, which may be used to periodically run automated tests. Jenkins is also preferable because it supports a variety of version managers (e.g. CVS or Git in addition to SVN). As for test automation, Jenkins is capable of running JUnit tests. The architecture of Jenkins also enables the extension of functionality and thus can be integrated with numerous other testing frameworks [1]. Another possible path of development worth mention is integration into the IBM Rational DOORS requirement management application. This step would fulfill the expectations laid out in the V-model. This solution could also help in managing the scope and costs of the project and in meeting business goals. With the test follow-up repertoire of DOORS, tools designed for a manual test follow-up environment may be linked to the requirements. By modifying the requirements, a predefined change proposal system may be used under given circumstances when the unit test verification is changed (see [3]). The efficiency of the unit test presented above may be examined by a software (Sonar program), which can provide us information on the coverage of the tested code, and the fail rate of the tests. While this information shows only tendencies, they can still serve as a general guide and provide confirmation on the efficiency of the unit testing.

## V. Conclusions

In this article we presented the first results of a PhD study connected to testing of safety critical medical devices. The article summarizes how a Hungarian manufacturer of medical devices actually uses a self-developed software tool for automated testing of a medical device. Testing is extremely critical in this environment, therefore we expected to find within the company testing principles, methods and tools above average. Our expectations were met: the company subject of the case study has a rather mature testing process in place, and invested in developing a testing tool to automate the unit testing process. A significant advantage of the DSS unit test currently in use is that GUI may be circumvented and so hardware operation may be checked without GUI. This enables the functional, isolated testing of individual hardware elements. Through choosing the right timer task, we might gain additional information on development deficiencies that are difficult or impossible to notice with high level tests (manual or system test), as the CMs are not visible in a 250ms breakdown. Besides, the unit test fits V-model requirement rules and thus the criteria in the standard are fulfilled. The specification based unit test of the V-model currently in use does not provide a global picture of the system. When making a system model, however, all input and output points in the communication between the interfaces are visible on the state chart. This makes designing and checking test cases easier. Another advantage of system model design is that it makes hidden inconsistencies in the specifications easier to identify. Thus, investigating whether implementation matches the functions described in the specification is more efficient. With the model based test generating tool as a DSS verification process, specification based testing may be designed on a global level. Using this method can also aid the developers, as development objectives in connection with the requirements are easier to interpret, just like test cases during test design. Medical devices have critical safety elements that affect not only safe operation but also the treatment of the patient. Linking the tests of these critical safety elements with risk estimation functions would thus be beneficial. Such estimation would make it possible to assess the severity level of test cases. However, as testing must be efficient, analysis of testing results must be made automatic. Some steps have been already made in this direction, but further research is needed. The article concludes by pointing out several possible directions for this research, based on the author's analysis of possibilities.